\documentclass[11pt,letterpaper]{JHEP3}
\pdfoutput=1
\usepackage{cite}

\usepackage{amsmath,amssymb,amsfonts}

\usepackage[english]{babel}
\usepackage{amssymb}
\usepackage{amsfonts}
\usepackage{amsmath}
\usepackage{graphicx}
\usepackage{caption,subfig}

\renewcommand\d{\partial}

\newcommand{\be}{\begin{equation}} \newcommand{\ee}{\end{equation}}
\newcommand{\bea}{\begin{eqnarray}} \newcommand{\eea}{\end{eqnarray}}
\newcommand{\beann}{\begin{eqnarray*}}  \newcommand{\eeann}{\end{eqnarray*}}
\newcommand{\bfig}{\begin{figure}} \newcommand{\efig}{\end{figure}}
\newcommand{\ba}{\begin{array}} \newcommand{\ea}{\end{array}}
\newcommand{\bcen}{\begin{center}} \newcommand{\ecen}{\end{center}}
\newcommand{\btab}{\begin{tabular}} \newcommand{\etab}{\end{tabular}}
\newcommand{\nn}{\nonumber}
\title{Asymptotically hyperbolic black holes in Ho\v rava gravity}

\author{
Stefan Janiszewski\footnotemark[1]
\\
Department of Physics, University of Washington, Seattle, WA
98195-1560}

\footnotetext[1]{E-mail: \email{stefanjj@u.washington.edu}}

\abstract{Solutions of Ho\v rava gravity that are asymptotically Lifshitz are explored. General near boundary expansions allow the calculation of the mass of these spacetimes via a Hamiltonian method. Both analytic and numeric solutions are studied which exhibit a causal boundary called the universal horizon, and are therefore black holes of the theory. The thermodynamics of an asymptotically Anti-de Sitter Ho\v rava black hole are verified.}

\begin{document}
\maketitle
\flushbottom

\section{Introduction}
Recently, it has been shown that the Lifshitz spacetime,
\begin{equation}\label{eq:lifshitzst}
ds^2= -\left(\frac{L}{r}\right)^{2z}dt^2+\left(\frac{L}{r}\right)^{2}dr^2+\left(\frac{L}{r}\right)^{2}d\vec{x}^2,
\end{equation}
is a vacuum solution to Ho\v rava gravity \cite{Griffin:2012qx,Janiszewski:2012nb}. Much as the maximally symmetric spacetimes of general relativity (GR) form a starting point to explore the landscape of solutions of that theory, Eq. \ref{eq:lifshitzst} implies that Ho\v rava gravity has a sector of vacuum solutions that asymptotically approach the novel structure of Lifshitz spacetimes. Further motivation comes from the arena of holography where the Anti-de Sitter spacetime, that is Eq. \ref{eq:lifshitzst} with $z=1$, is argued to be dual to relativistic quantum field theories \cite{Maldacena:1997re,Witten:1998qj,Gubser:1998bc}. Indeed, for $z\ne 1$ it has been shown that Lifshitz spacetimes are holographically dual to non-relativistic field theories \cite{Kachru:2008yh,Balasubramanian:2010uk}.

Based on these considerations, after introducing Ho\v rava gravity in Section \ref{Horavagravity}, we explore the space of solutions that are asymptotically Lifshitz, Eq. \ref{eq:lifshitzst}. This is done in Section \ref{sec:asymp} by expanding the equations of motion about asymptotic infinity and solving order by order. The resulting power series solution for the spacetime is quantified by some number of free constants. For example, in GR the Schwarzschild solution can be found this way by expanding about asymptotically flat space: the free constant in the power series solution corresponds to its mass. 

The main result of this paper is a presentation of asymptotically Lifshitz black holes in Ho\v rava gravity. The preliminary discussion in Section \ref{horizons} is required to justify the existence of such objects, as the causal structure of the theory is very different from GR. Section \ref{sec:num} exhibits numerical solutions that have sensible horizons and are rightly called black holes. In Section \ref{sec:anal}, an analytic black hole found by examining the asymptotic expansion is dissected. Importantly, it can be assigned mass, temperature, and entropy that obey a consistent first law of thermodynamics.

\section{Ho\v rava gravity}\label{Horavagravity}
Ho\v rava gravity is an alternate theory of gravity \cite{Horava:2009uw}. Like general relativity, its low energy behavior can be expressed in the context of geometry. In GR a dynamic spacetime manifold encodes the influence of gravity: the manifold responds to the presence of mass and energy, while simultaneously the geometry dictates the evolution of said matter. A hallmark of GR, capturing its geometric nature, is the coordinate invariance of observables. This is captured in the physical description by demanding that diffeomorphisms, that is, local changes of coordinates on the manifold, are a gauge symmetry of the theory. 

Ho\v rava gravity shares much of this geometric flavor, but, unlike GR, in addition to a dynamic spacetime there is a fundamentally special foliation, $\Sigma_t$, of the manifold. Ho\v rava gravity has a preferred notion of time which is geometrically captured by the leaves of a co-dimension one foliation: these slices consist of simultaneous events. The foliation structure breaks the general covariance enjoyed by just the manifold. The preferred global time of Ho\v rava gravity means that Lorentzian coordinate changes that mix spatial directions into a new time coordinate are no longer allowed. These would alter the notion of which events are simultaneous and violate the preferred foliation of the manifold. The temporal coordinate can still be reparametrized: sending $t\to \tilde{t}(t)$ for an arbitrary monotonic function $\tilde{t}$ preserves the simultaneity of events, and hence the foliation structure. Spatial diffeomorphisms that change coordinates on a leaf are still allowed, and indeed can even be time dependent. These transformations, spatial diffeomorphisms $x_I\to \tilde{x}_I(t,x_I)$\footnote{We use indices $I,J,\dots$ to denote the $D$ spatial coordinates of the geometry, while $M,N,\dots$ will cover all $D+1$ coordinates of the manifold.} and time reparmetrizations $t\to \tilde{t}(t)$, are collectively called foliation preserving diffeomorphisms (FDiffs), and are a gauge symmetry of Ho\v rava gravity.

The degrees of freedom of Ho\v rava gravity can be understood in this low energy geometric picture. A spacetime manifold in coordinates adapted to the foliation will have a metric of the ADM form:
\begin{equation}\label{ADM}
 g_{MN}dx^M dx^N=-N^2dt^2+G_{IJ}\left(dx^I+N^Idt\right)\left(dx^J+N^Jdt\right).
\end{equation}
Here $G_{IJ}$ is the spatial metric on the leaves of the foliation at a constant $t$, $N$ is the lapse function, and $N^I$ is the shift, a spatial vector. The low energy action of Ho\v rava gravity is expressed in terms of these fields as \cite{Griffin:2012qx}\footnote{Note that our coupling constants are not the same as \cite{Griffin:2012qx}, despite identical names.}:
\begin{equation}
 \label{haction}
S_H=\frac{1}{16\pi G_H}\int dt d^Dx N\sqrt{G}\left(K_{IJ}K^{IJ}-(1+\lambda)K^2+(1+\beta)(R-2\Lambda)+\alpha \frac{\nabla_I N \nabla^I N}{N^2}\right),
\end{equation}
where $K_{IJ}\equiv\frac{1}{2N}(\d_t G_{IJ}-\nabla_I N_J-\nabla_J N_I)$ is the extrinsic curvature of the leaves of the foliation, $R$ is the Ricci scalar of the spatial metric $G_{IJ}$, $G$ is its determinant, and $\Lambda$ is the cosmological constant. The dimensionless couplings $(\alpha,\beta,\lambda)$ are new parameters allowed by the less restrictive FDiff symmetry group, whereas the action of GR written in terms of these fields would require them to vanish. The constant $G_H$ has mass dimension $[G_H]=1-D$ and sets the Planck mass.

Ho\v rava gravity departs much more drastically from GR in its high energy behavior. Although this paper will not be concerned with this regime a brief discussion is pertinent. Because of the fundamental foliation of spacetime a preferred notion of time exists in Ho\v rava gravity and Lorentz symmetry is broken. This allows temporal and spatial coordinates to have different mass dimensions and is captured by the dynamical critical exponent $z_H$: $[x_I]=-1$ while $[t]=-z_H$, implying $[G_H]=z_H-D$. An interesting case is near a UV fixed point with $z_H=D$. The UV action will be dominated by terms with $2D$ spatial derivatives, while there will be the same kinetic terms involving $K_{IJ}$ as in Eq. \ref{haction}, which contain only two time derivatives. This appears to lead to a unitary, power counting renormalizable theory, as evidenced by the fact that now $[G_H]=0$. The low energy action Eq. \ref{haction} with an effective $z_{IR}=1$ would then be the flow from the UV fixed point under relevant deformations.

\subsection{Khronon formalism}\label{khronon}

The geometric nature of low energy Ho\v rava gravity makes it possible to cast the action Eq. \ref{haction} in a generally covariant form. Intuitively, this is done by capturing the dynamics of the foliation by a scalar field coupled nontrivally to Einstein gravity. This scalar field $\phi$ is called the khronon \cite{Germani:2009yt,Blas:2010hb} and its level sets define the leaves of the foliation. To have the desired time reparametrization of Ho\v rava gravity $t\to \tilde{t}(t)$ the khronon needs the field space reparametrization invariance $\phi\to f(\phi)$ for arbitrary monotonic $f$. This can be made explicit in the action by using the invariant derived quantity:
\begin{equation}
u_M\equiv\frac{-\d_M \phi}{ \sqrt{-g^{NP}\d_N \phi \d_P \phi}},
\end{equation}
which is the unit time-like vector normal to the hypersurfaces determined by the leaves of the foliation. The most general covariant action involving two derivatives of the normal vector $u_M$ is related to Einstein-aether theory \cite{Jacobson:2000xp}, and given by:
\begin{equation}\label{kaction}
 S_K=\frac{1}{16 \pi G_K}\int \sqrt{-g}\left(\mathcal{R}-2\Lambda+c_4 u^M\nabla_M u^N u^P\nabla_P u_N-c_2 \left(\nabla_M u^M\right)^2-c_3\nabla_M u^N\nabla_N u^M\right),
\end{equation}
where $g$ is the determinant of the full spacetime metric, $\mathcal{R}$ is its Ricci scalar, the $c_i$ are coupling constants, and the hypersurface orthogonality of $u_M$ has been used. 

To make contact with Ho\v rava gravity one chooses the scalar field $\phi$ to be the time coordinate $t$, breaking the general covariance of the khronon-metric theory. This gives only a temporal component to the normal vector, $u_M=-N\delta^t_M$, while the spacetime metric $g_{MN}$ is decomposed in the ADM form Eq. \ref{ADM} with respect to the time $t\equiv\phi$. The khronon action Eq. \ref{kaction} then becomes to the low energy Ho\v rava action Eq. \ref{haction} upon making the identification of constants \cite{Barausse:2011pu}:
\begin{equation}\label{ccs}
 \frac{G_H}{G_K}=1+\beta=\frac{1}{1-c_3},\quad 1+\lambda=\frac{1+c_2}{1-c_3},\quad \alpha = \frac{c_4}{1-c_3}.
\end{equation}
By examining the weak-field, slow-motion limit of the action Eq. \ref{kaction} it is seen that the effective Newton's constant is \cite{Carroll:2004ai,Barausse:2011pu}:
\begin{equation}\label{newtonconst}
 G_N\equiv \frac{1-c_3}{1-\frac{c_4}{2}}G_H.
\end{equation}
Lastly, the equations of motion following from Eq. \ref{kaction} can be linearized around the flat background solution to determine the speed of the modes. The low momentum dispersion relations determine the wave speeds squared to be \cite{Jacobson:2004ts,Griffin:2012qx}:
\begin{equation}\label{speed}
 s^2_2=\frac{1}{1-c_3}, \quad s^2_0=\frac{(c_2+c_3)(D-1-c_4)}{c_4(1-c_3)(D-1+Dc_2+c_3)},
\end{equation}
for the spin two modes of the metric and the spin zero mode of the foliation, respectively.

A powerful use of the khronon formalism is in the probe regime where the $c_i$ are parametrically small. In this case the backreaction of the khronon on the geometry can be ignored and any solution of vacuum GR descends to a solution of Ho\v rava gravity. In the probe limit the action Eq. \ref{kaction} is just that of GR, so the metric equations of motion are clearly satisfied. The equation of motion for the khronon $\phi$ can then be solved on the fixed background given by the metric solution. By making the Lorentzian coordinate change $t\to\hat{t}\equiv\phi(t,x_I)$ the resulting lapse $N$, shift $N_I$, and spatial metric $G_{IJ}$ of the decomposition Eq. \ref{ADM} are now a solution to Ho\v rava gravity. In the probe limit the khronon does not influence the geometry of the manifold, it merely imprints the preferred notion of time via its level sets.

From the identification Eq. \ref{ccs} the probe limit of the khronon formalism gives the Ho\v rava coupling constants:
\begin{equation}
 \beta\approx c_3\ll 1,\quad \lambda\approx c_2+c_3\ll 1, \quad\alpha \approx c_4\ll 1,
\end{equation}
while the speeds of the modes Eq. \ref{speed} reduce to:
\begin{equation}
 s_2^2\approx 1,\quad s_0^2\approx \frac{c_2+c_3}{c_4}=\frac{\lambda}{\alpha}.
\end{equation}
Interestingly, even in the probe limit the scalar mode can have arbitrary sound speed. In taking the limit this physical speed should be held fixed; that is, the ratio $\lambda/\alpha$ is held fixed while both constants are taken to zero. This technique will be used in Section \ref{sec:num} to compute numerical profiles for the khronon in the background of an Anti-de Sitter-Schwarzschild black hole.

\subsection{Horizons and thermodynamics}\label{horizons}
The lack of Lorentz invariance makes causality a subtler notion in Ho\v rava gravity than it is in GR. In Einstein gravity the causal structure of a solution is most apparent when brought into Penrose form. Light cones form forty five degree diagonal lines and define the invariant notion of whether one event is space-like, null, or time-like separated from another. This nature of separation between two events and the domain of influence for a given region are easily deduced from the Penrose diagram. From this construction event horizons are identified as the null boundary that separates the domain of dependence of future infinity from the rest of the manifold.

In Ho\v rava gravity a light cone is not a limiting object, and the notion of causality is maintained by the existence of a preferred global time instead. Indeed, as seen in the previous section, fields in Ho\v rava gravity can have arbitrarily fast propagation speed, there is no limiting role of the speed of light that is fundamental to GR. Despite this, a mode traveling with any speed in the preferred frame can only propagate forward in global time. The leaves of the foliation labeled by the preferred time define the invariant notion of whether one event is before, simultaneous with, or after another.

\begin{figure}[!htb]
\centering
\subfloat[][]{\includegraphics[width=5.3cm]{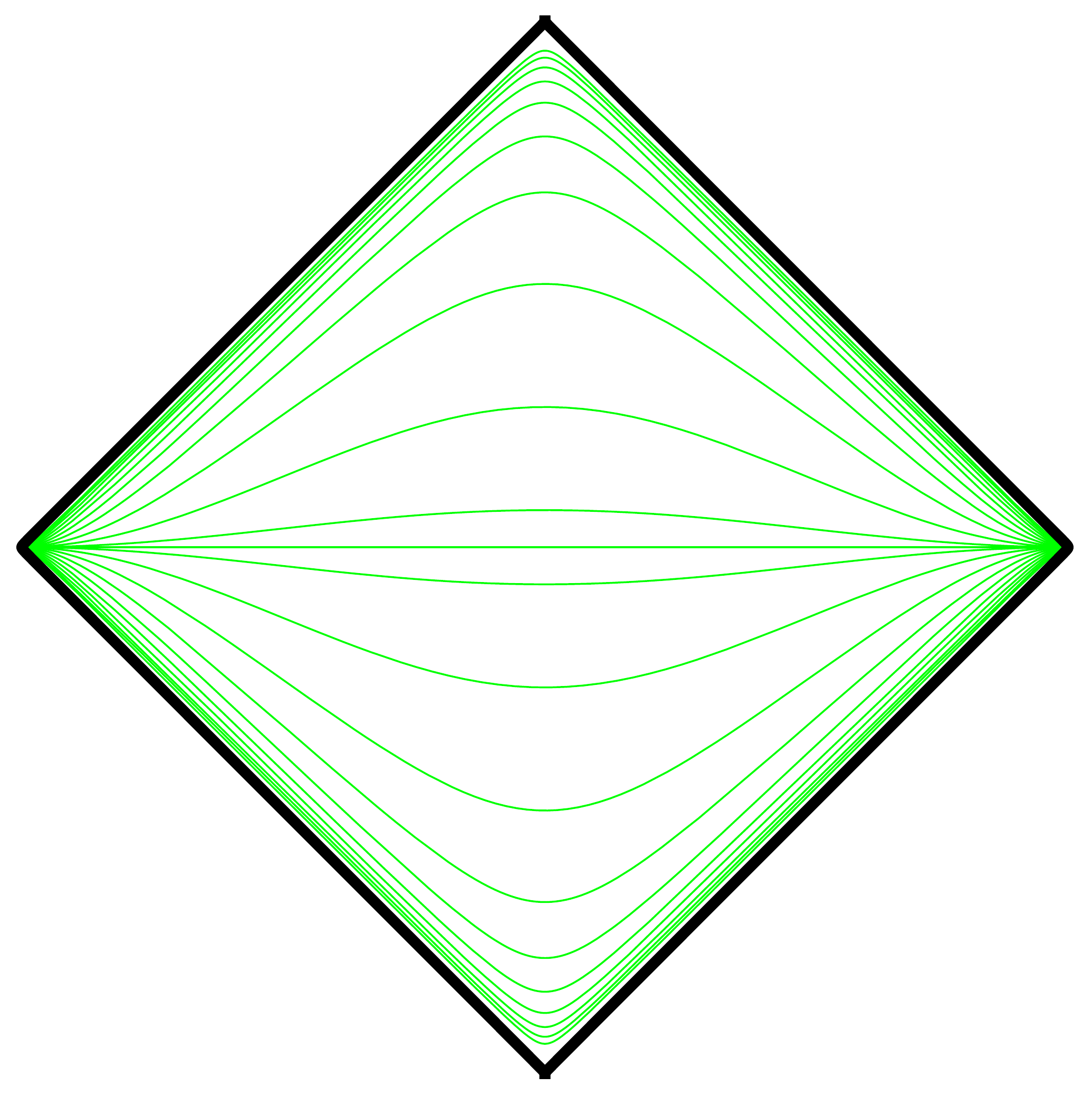}}%
\quad
\subfloat[][]{\includegraphics[width=8cm]{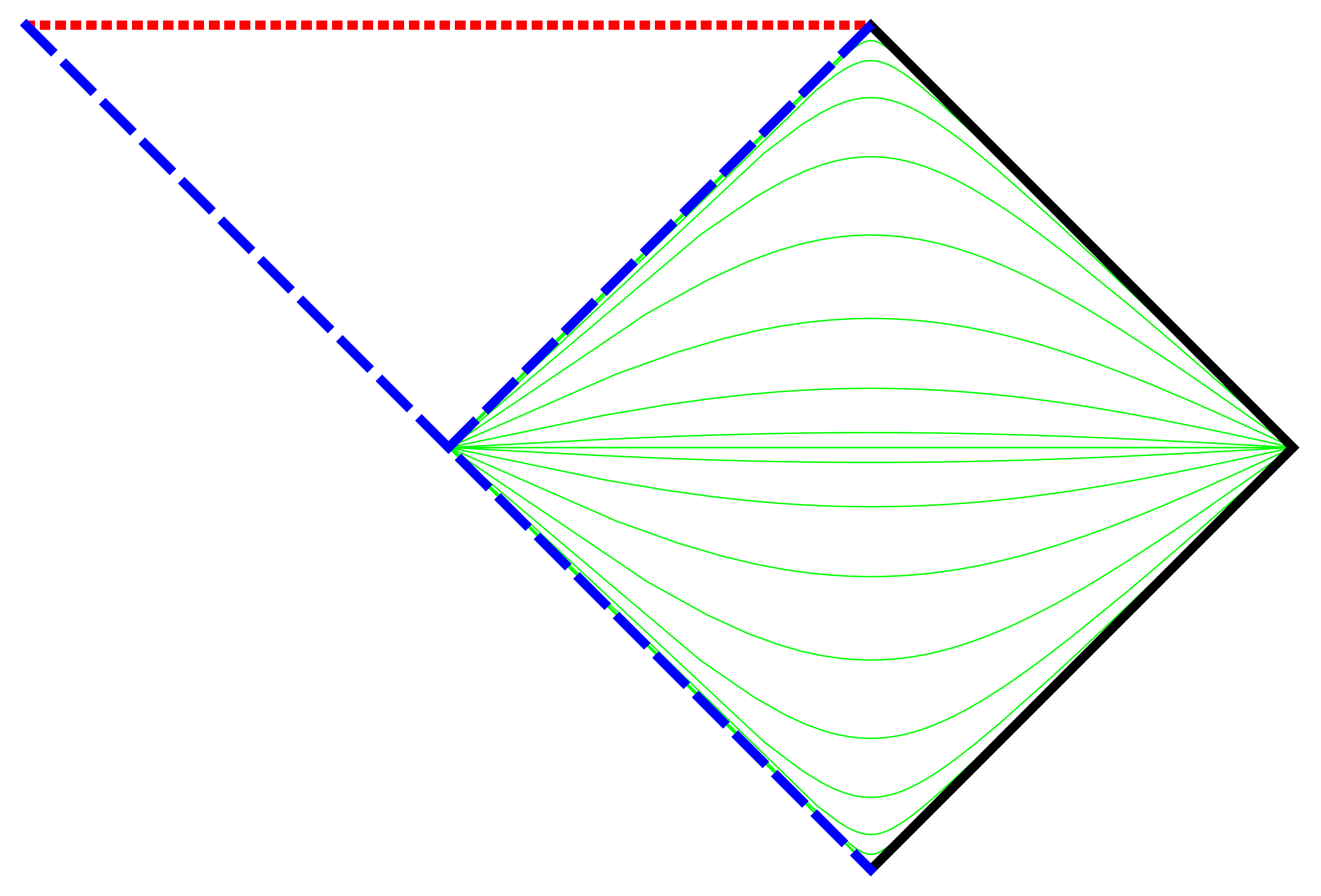}}%
\caption{a) The Penrose diagram for Minkowski space. The thin green lines are slices of constant time, while $t=\pm\infty$ is null infinity in solid black (color online). b) The Penrose diagram for a Schwarzschild black hole. The thin green lines are of constant Schwarzschild time, $t=\pm\infty$ is partially null infinity in solid black, and partially the event horizon in dashed blue. The singularity at $r=0$ is in dotted red.}
\label{fig:penschwr}
\end{figure}

Likewise, the foliation by a preferred time can define the notion of causal boundaries and horizons in Ho\v rava gravity. Simplistically, the leaf of the foliation labeled by $t=\infty$ forms a causal boundary of the manifold. Only events on leaves labeled by earlier times can possibly influence this boundary. Naively, this leaf may be expected to simply form the boundary of the manifold defined as future infinity, much as $t=\infty$ is future null infinity in Minkowski space. More interestingly, this leaf can ``bend down'' as it foliates the manifold and may never penetrate some region. This is analogous to the asymptotically flat Schwarzschild black hole in Schwarzschild time: the line $t=\infty$ is partially asymptotic null infinity, but partially the the event horizon at $r=r_h$, see Figure \ref{fig:penschwr}.

It is important to note that in GR, the fact that leaves of Schwarzschild time converge at the event horizon is not an invariant statement about the geometry of the spacetime, different coordinates can have a foliation that does nothing special at $r_h$. To identify the leaf $t=\infty$ as a causal boundary of the black hole spacetime in GR we also need to use the fact that it is a null surface. Only then is it concluded that nothing inside the radius $r_h$ can influence events outside: that is, the leaf $t=\infty$ does form a causal boundary of the foliation. 

\begin{figure}[!htb]
\centering
\includegraphics[width=0.42\textwidth]{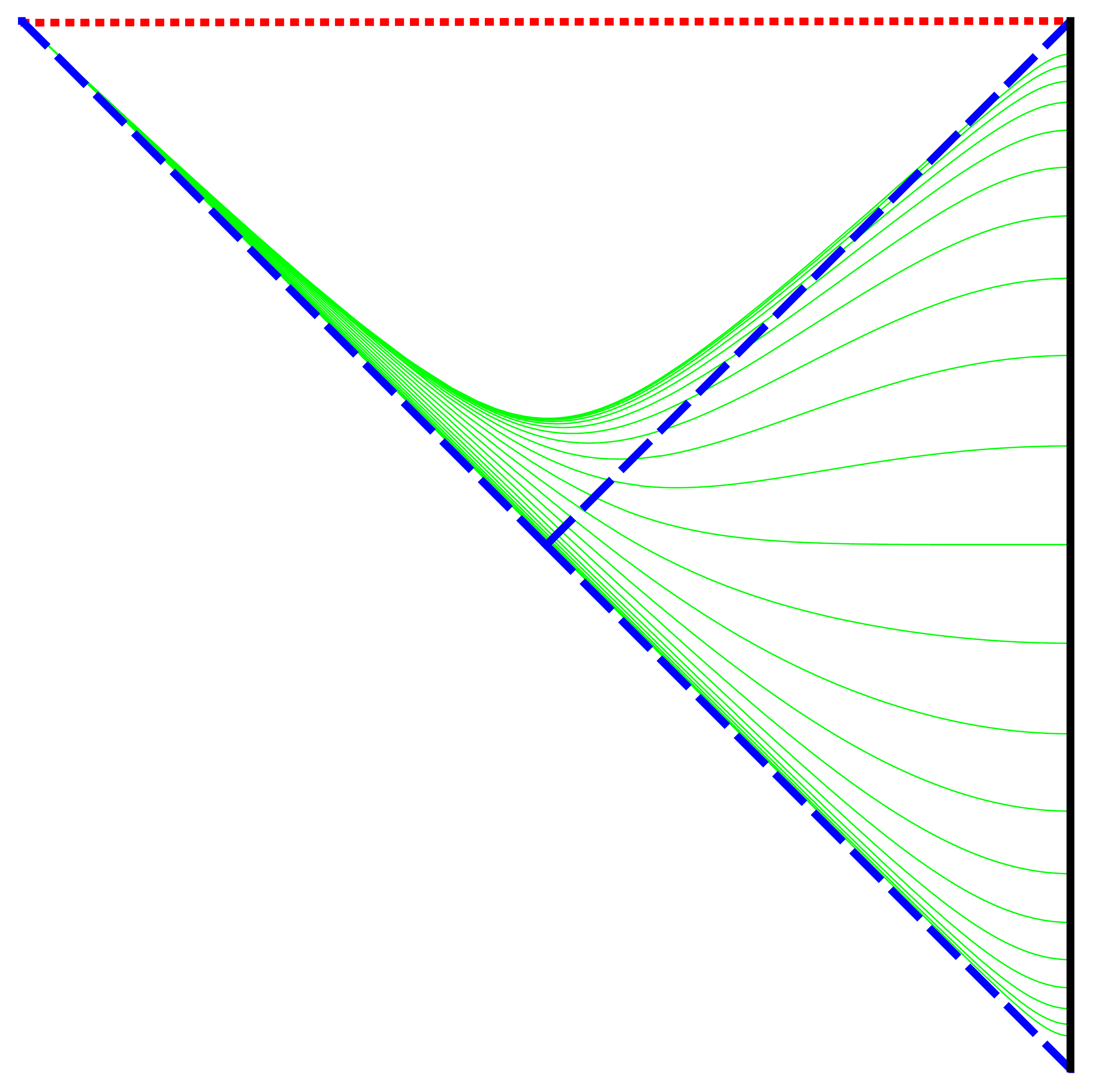}
\caption{The Penrose diagram of a Ho\v rava black hole. The singularity is in dotted red at the top, the boundary is in solid black on the right, the metric horizon is in dashed blue. The leaves of the foliation are in thin green and pile up at the universal horizon.}
\label{fig:pensads}
\end{figure}

The case in Ho\v rava gravity is somewhat simpler in this respect. Causality is encoded in the foliation itself, not any light cone structure. The convergence of its leaves is an invariant statement, and such a region is a causal boundary. Such a slice (that isn't asymptotic infinity) is called a ``universal horizon'' \cite{Eling:2006ec,Barausse:2011pu,Blas:2011ni}: an event behind it cannot influence events outside as they are all at ``earlier'' times as measured by the preferred foliation. Solutions to Ho\v rava gravity with universal horizons will be called black holes. An example, which is asymptotically Anti-de Sitter space and constructed in the following sections, is shown in Figure \ref{fig:pensads}.

The existence of causal horizons in Ho\v rava gravity begs the question of whether they obey a thermodynamic description as do their counterparts in GR \cite{Blas:2011ni,Berglund:2012bu,Berglund:2012fk}. Motivation for such a description follows from arguments analogous to Bekenstein's original proposals \cite{PhysRevD.7.2333}: a causal horizon must have intrinsic entropy if the second law is not to be violated when exterior entropy falls in. Consistency with a first law then implies a temperature of the horizon. For Ho\v rava gravity, additional motivation comes from its claim as a UV complete quantum theory of gravity \cite{Blas:2011ni}. From a microscopic description, the entropy of a macroscopic system is a measure of the number of fundamental degrees of freedom contained. The macroscopic second law is then a reflection of the unitarity of the microscopic theory. If Ho\v rava gravity is truly a sensible quantum theory of gravity macroscopic systems such as black holes defined by their universal horizons must obey a second law of thermodynamics.

Holography \cite{'tHooft:1993gx,Susskind:1994vu} and AdS/CFT \cite{Maldacena:1997re,Witten:1998qj,Gubser:1998bc} grew out of understanding black hole thermodynamics in GR, and from combining principles of quantum mechanics and black hole entropy. If these same arguments can be applied to the universal horizons of Ho\v rava gravity then the notion of holography naturally extends to a much larger class of theories, most notably those of an intrinsic non-relativistic nature \cite{Horava:2009uw,Berglund:2012bu,Berglund:2012fk,Janiszewski:2012nb,Janiszewski:2012nf,Griffin:2012qx}.

\section{Asymptotic solutions}\label{sec:asymp}

Understanding the asymptotic structure of gravitational theories has proven to be a fruitful endeavor. Various ``No Hair'' theorems have used these techniques to determine that black holes in GR are labeled by a small number of parameters. In the context of Einstein-aether theory and Ho\v rava gravity this has been done for asymptotically flat solutions \cite{Eling:2006ec,Barausse:2011pu,Berglund:2012bu}. Generically, static spherically-symmetric solutions have three defining parameters. The requirement of asymptotic flatness reduces this number to two. The solutions are reduced to a one parameter family, the mass, upon requiring regularity at the spin-zero sound horizon, that is, the trapped surface for waves of the speed $s_0$ in Eq. \ref{speed} \cite{Eling:2006ec,Barausse:2011pu}.

\subsection{Hyperbolic spacetimes}
The focus of this paper is understanding the space of solutions to Ho\v rava gravity that have an asymptotically Lifshitz metric:
\begin{equation}\label{eq:lif}
 \lim_{r\to 0} ds^2\approx -\left(\frac{L}{r}\right)^{2z}dt^2+\left(\frac{L}{r}\right)^{2}dr^2+\left(\frac{L}{r}\right)^{2}d\vec{x}^2,
\end{equation}
where $z$ is the dynamical critical exponent that controls the anisotropy of time versus space, and $L$ is a length giving the scale of the curvature. As the spatial metric given by slices of constant $t$ has uniform negative curvature, Eq. \ref{eq:lif} will be referred to as a hyperbolic spacetime. Metrics of this form are important in the arena of holography. Allowing $z=1$, one obtains the metric of Anti-de Sitter (AdS) space in Poincar\'e coordinates. Such boundary conditions are crucial from the viewpoint of standard AdS/CFT \cite{Maldacena:1997re,Witten:1998qj,Gubser:1998bc}. For $z\ne1$ the metric Eq. \ref{eq:lif} exhibits an anisotropic scaling symmetry of time relative to space. General relativity on such backgrounds has been argued to be dual to non-relativistic field theories \cite{Kachru:2008yh,Balasubramanian:2010uk}. More pertinently, although GR requires additional matter to support this geometry, such a metric is a vacuum solution to Ho\v rava gravity \cite{Griffin:2012qx,Janiszewski:2012nb}. This leads to speculation that the Lifshitz metric may play the fundamental role in a holographic duality involving Ho\v rava gravity that AdS space plays in traditional holography.

The Poincar\'e-like coordinates of Eq. \ref{eq:lif} are a natural choice to work in as they most simply lead to the definition of the boundary at $r=0$ via an anisotropic conformal transformation \cite{Horava:2009vy}. However, in order to fully explore the interior geometry of an asymptotically hyperbolic spacetime it is important to use coordinates that are nowhere singular. From experience with black holes in Minkowski and AdS spaces, the partial-null Eddington-Finkelstein-like (EF) coordinates are a better choice: they are not singular at the metric horizon, unlike Schwarzschild and Poincar\'e coordinates. To obtain the Lifshitz metric Eq. \ref{eq:lif} in EF coordinates one first defines the radial tortoise coordinate $r_*\equiv1/z(r/L)^z$. From this the EF time is defined as the null coordinate $v\equiv t+r_*$, and the metric becomes:
\begin{equation}\label{EFmetric}
 ds^2=-\frac{dv^2}{r^{2z}}+2\frac{dvdr}{r^{z+1}}+\frac{d\vec{x}^2}{r^2},
\end{equation}
where units have been chosen such that $L=1$.

In Ho\v rava gravity, in addition to the asymptotic behavior of the metric, the boundary conditions of the foliation must be specified. For holography it is natural to use a time coordinate that is asymptotically the Poincar\'e time of Eq. \ref{eq:lif}: this time appears as the conformal boundary time and would correspond to a global time coordinate of the dual field theory on a flat background. In the khronon language this requires $\phi|_{r\to 0}=t=v-r_*$, or in terms of the foliation normal vector in EF coordinates:
\begin{equation}\label{EFaether}
 u_M=\left(-\frac{1}{r^z},\frac{1}{r},\vec{0}\right).
\end{equation}

Respecting the stationarity and planar symmetry properties of the Lifshitz metric Eq. \ref{eq:lif}, the general ansatz of the full metric in EF coordinates is:
\begin{equation}
 \label{ansatzm}ds^2=-e(r)dv^2+2f(r)dvdr+\frac{d\vec{x}^2}{r^2}.
\end{equation}
The general foliation normal vector respecting these symmetries is:
\begin{equation}\label{ansatza}
 u_M=\left(-\frac{a(r)^2e(r)+f(r)^2}{2a(r)f(r)},a(r),\vec{0}\right),
\end{equation}
where the unit norm constraint has been imposed. The required asymptotic behavior to reproduce a hyperbolic spacetime, Eq. \ref{EFmetric} and Eq. \ref{EFaether}, gives the leading behavior as $r\to0$ of the free functions:
\begin{equation}\label{bc}
 e(r)\sim \frac{1}{r^{2z}},\quad f(r)\sim\frac{1}{r^{z+1}},\quad a(r)\sim \frac{1}{r}.
\end{equation}

The remainder of this paper will be restricted to $D=3$, that is, a four dimensional spacetime. This allows explicit expressions to be written down simply, although generalization is straightforward. A useful trick to simplify the action Eq. \ref{kaction} is to note that a hypersurface orthogonal vector such as $u_M$ has vanishing curl:
\begin{equation}
u_{[N}\nabla_P u_{Q]}=0.
\end{equation}
Therefore, in four dimensions, for $\omega_M\equiv \epsilon_{MNPQ}u^N\nabla^Pu^Q=0$, we can write:
\begin{equation}
 (u^M \nabla_M u_N)^2=(u^M \nabla_M u_N)^2-\omega_M\omega^M=-\frac{1}{2}F_{MN}F^{MN},
\end{equation}
for the ``field strength'' $F_{MN}\equiv \d_M u_N-\d_N u_M$. This simplifies the $c_4$ term in the action as it removes any factors of the connection in the derivative. The task is to now examine the equations of motion coming from the action Eq. \ref{kaction} with the ansatz Eq. \ref{ansatzm} for the metric and Eq. \ref{ansatza} for the foliation normal vector. Plugging in a series solution for the functions $(e,f,a)$, with leading behavior given by Eq. \ref{bc}, the equations of motion can be solved order by order as $r\to0$. This gives evidence to the number of free parameters that classify a general solution by determining the number of free coefficients in the series expansions of $(e,f,a)$. The full equations of motion are cumbersome and not very enlightening, and will not be shown here. To zeroth order the solution is simply the Lifshitz background Eq. \ref{EFmetric}, requiring \cite{Griffin:2012qx}:
\begin{equation}\label{asymplif}
 c_4 = \frac{z-1}{z},\quad \Lambda=-\frac{(D-2+z)(D-1+z)}{2},
\end{equation}
that is, the dynamical critical exponent $z$ is determined by the coupling $c_4$, and the cosmological constant is in turn fixed. To higher order there is a drastic difference whether $z=1$ or not.

\subsubsection{$z=1$}
For $z=1$ ($c_4=0$) the metric is asymptotically Anti-de Sitter space. The general asymptotic series expansion has coefficients proportional to $(z-1)$, therefore this case must be treated separately. Examining the order by order expansion of the equations of motion as $r\to0$ it appears the series for $e$ and $f$ truncates. To order $r^{30}$ the functions are determined to be:
\begin{eqnarray}\label{zeq1exp}
 a(r)&=&\frac{1}{r}+C_a r^2+\frac{1}{2}\left((c_3+1)C_a^2+(c_3-1)C_aC_e+\frac{c_3}{4}C_e^2\right)r^5+\cdots,\\
e(r)&=&\frac{1}{r^2}+C_e r-\frac{1}{4}c_3(C_e+2C_a)^2r^4,\\
f(r)&=&\frac{1}{r^2},
\end{eqnarray}
where $C_e$ and $C_a$ are constants, and only the first three terms of $a(r)$ are shown, although importantly no more free parameters appear. The truncation of an asymptotic series solution is often the signature of an analytic solution, which will indeed be constructed in Section \ref{sec:anal}.

\subsubsection{$z\ne1$}
For $z\ne 1$ the asymptotic series analysis is more subtle. Stripping off the leading boundary behavior Eq. \ref{bc} and expanding the free functions in a series solution, as above, gives:
\begin{eqnarray}\label{zneq1exp}
 a(r)&=&\frac{1}{r}\left(1-\frac{1}{2}C_er^{z+2}+\frac{(2z^2+2z-1)C_e^2}{4z(z+1)}r^{2(z+2)}-\frac{(9z^2+9z-8)C_e^3}{16z(z+1)}r^{3(z+2)}+\cdots\right)\nn \\
e(r)&=&\frac{1}{r^{2z}}\left(1+C_er^{z+2}+\frac{(z-1)(z+2)C_e^3}{24z(z+1)}r^{3(z+2)}-\frac{(z-1)(z+2)C_e^4}{24z(z+1)}r^{4(z+2)}+\cdots\right)\nn \\
f(r)&=&\label{eq:neone}\frac{1}{r^{z+1}}\left(1+\frac{(z-1)(z+2)C_e^2}{8z(z+1)}r^{2(z+2)}-\frac{(z-1)(z+2)C_e^3}{6z(z+1)}r^{3(2+z)}+\cdots\right).
\end{eqnarray}
Importantly, up to order $r^{5(z+2)}$, the expansion is seen to contain only one free coefficient, $C_a$, contrary to general expectations \cite{Eling:2006ec}. Also evident is the fact that, apart from the leading singular factor, the functions $(a,e,f)$ are only functions of $r^{z+2}$. 

To find the missing constant it helps to recognize when the above procedure fails. It has been assumed that the free functions can be expanded as their leading singular behavior times an analytic function. Whether this analyticity is justified depends on what the subleading characteristic exponents are. These can be determined by making the ansatz:
\begin{equation}\label{eq:sublead}
 e(r)= \frac{1}{r^{2z}},\quad f(r)=\frac{1}{r^{z+1}},\quad a(r)= \frac{1}{r}\left(1+a_\Delta r^\Delta\right),
\end{equation}
and using the equations of motion for $r\to0$ to determine the allowed powers $\Delta$. This procedure yields:
\begin{equation}\label{delta}
 \Delta_\pm=\frac{1}{2}\left(z+2\pm\sqrt{(z+2)^2-\frac{8(1-c_3)(z-1)}{c_2+c_3}}\right).
\end{equation}
The only requirement on $\Delta$ is that it is non-negative in order to maintain the desired boundary behavior of $a(r)$. Generically, $\Delta_\pm$ is non-integer and therefore the ansatz of analyticity is not justified. Indeed, $z=1$ is a special case for which $\Delta_\pm=(0,3)$.

In general for $\Delta$ to be an integer requires a specific choice of couplings. For concreteness take the case of $z=2$. Then for the choice $c_2=(1-3c_3)/2$ it is seen that $\Delta_{\pm}=2$. In this case the asymptotic series expansion of the functions is:
\begin{eqnarray}\label{z2delta2exp}
 a(r)&=&\frac{1}{r}\left(1+C_ar^2+\frac{1}{2}(C_a^2-C_e)r^4+(C_a(C_a^2-C_e)-C_a^3)r^6+\cdots\right)\nn\\
e(r)&=&\frac{1}{r^4}\left(1+C_er^4-\frac{1}{18}\left(C_a^2+\frac{1}{2}(C_e-C_a^2)\right)^2\left(2(C_a^2-C_e)+C_a^2(3c_3-4)\right)r^{12}+\cdots\right)\nn \\
f(r)&=&\frac{1}{r^3}\left(1+\frac{1}{3}\left(C_a^2+\frac{1}{2}(C_e-C_a^2)\right)^2r^8\right.  \nn\\&+&\left. \frac{1}{9}\left(C_a^2+\frac{1}{2}(C_e-C_a^2)\right)^2\left(4(C_a^2-C_e)+C_a^2(c_3-9)\right)r^{12}+\cdots\right).
\end{eqnarray}
Up to $\mathcal{O}(r^{12})$ the expansion is seen to have two free parameters, $C_a$ and $C_e$, and reduces to Eq. \ref{eq:neone} for $C_a=0$. Importantly, as evident in the general case from Eq. \ref{eq:sublead}, only the subleading behavior of the function $a(r)$ is modified; generically the subleading piece of $e(r)$ goes as $r^{2-z}$, while that of $f(r)$ goes as $r^{z+3}$.

Thus for any $z$ it is found that planar symmetric stationary solutions of Ho\v rava gravity are determined by two constants: $C_e$, the coefficient of $r^{2-z}$ in the expansion of $e(r)$, and $C_a$, the coefficient of $r^{\Delta-1}$ in the expansion of $a(r)$. The demand for an asymptotically hyperbolic spacetime has reduced the three dimensional parameter space of solutions to these two. As in \cite{Eling:2006ec,Barausse:2011pu}, this parameter space can be argued to reduce to one constant for physically acceptable solutions. The key point is that although desired boundary conditions have been imposed at asymptotic infinity, there is no guarantee that the solutions are non-singular in the interior (disregarding singularities hidden by horizons). An important place to demand regularity of the solution is the spin-zero sound horizon, that is, the trapped surface for waves of the speed $s_0$ in Eq. \ref{speed}. This is physically reasonable as it is expected to be true for solutions that describe the late stages of gravitational collapse, as is argued in general relativity for the regularity of the metric horizon. This requirement of regularity reduces the two parameters describing an asymptotically hyperbolic Ho\v rava solution to one, the mass. 

\subsection{Spacetime mass}\label{sec:mass}
An important use of the near boundary asymptotic expansion of solutions is in the determination of the mass of a spacetime. Ho\v rava gravity has a preferred notion of time due to its foliation $\Sigma_t$. This leads us to define the mass of the spacetime as the on-shell Hamiltonian with respect to the preferred slicing, following \cite{Hawking:1995fd}. The first step in this process is to include the Gibbons-Hawking term in the action in order to make the variational problem well posed. This is accomplished by defining the total action:
\begin{equation}
 S_{T}\equiv \frac{1}{16\pi G_K}\int_\mathcal{M}\sqrt{-g}\mathcal{L}_K +\frac{1}{8\pi G_K}\int_{\d \mathcal{M}}\sqrt{|h|}\mathcal{K},
\end{equation}
where: $\mathcal{L}_K$ is determined from the khronon action Eq. \ref{kaction}; the first integral is over the manifold $\mathcal{M}$, while the second is the Gibbons-Hawking term over its boundary $\d \mathcal{M}$; and $\mathcal{K}$ is the trace of the extrinsic curvature of the boundary, while $h$ is the determinant of its induced metric.

When the Ricci scalar in $\mathcal{L}_K$ is decomposed with respect to the ADM variables of the foliation there arise two total derivatives:
\begin{equation}\label{bts}
 -\nabla_N\left(u^N\nabla_Mu^M\right)+\nabla_N\left(u^M\nabla_M u^N\right).
\end{equation}
These lead to additional boundary terms, and the structure of $\d\mathcal{M}$ deserves illucidation. For the asymptotically hyperbolic spacetimes of concern there is a time-like boundary at $r=0$. This has the outward space-like normal $s^N$, and we will assume that $s^Nu_N=0$ at $r=0$. Therefore the first boundary term arising from Eq. \ref{bts} does not contribute here. The second term combines with the Gibbons-Hawking term to give the extrinsic curvature of the two-surface of constant $t$ at $r=0$ as embedded in the preferred foliation, denoted $\phantom{}^2\mathcal{K}$.

The next components of $\d\mathcal{M}$ are the past and future space-like boundaries of the foliation, given schematically by $t=\pm \infty$. These surfaces have normal $u^N$, and therefore the boundary contribution from the second term of Eq. \ref{bts} vanishes due to the unit norm of $u^N$. The first term can then be seen to completely cancel the Gibbons-Hawking term, leading to no net boundary contributions from past and future infinity. 

The final boundary possible is that of the universal horizon for black hole spacetimes. As discussed in the introduction, and explored further later, this is a surface that the asymptotic foliation $\Sigma_t$ does not penetrate, and is the causal boundary of the spacetime. To understand its boundary contributions to the Hamiltonian it pays to be more precise. Consider a spacetime with a Killing vector $\chi^N$ that is asymptotically time-like. As one travels inwards along the leaves of $\Sigma_t$ the product $\chi^Nu_N$, which is initially negative, can approach 0 at some value of $r=r_h$ \cite{Berglund:2012bu}. The Killing vector is therefore tangent to this surface. Causal evolution in the direction of increasing global time $t$ is therefore necessarily toward the center of the spacetime, and can never reach the asymptotic boundary at $r=0$. Thus this surface at a constant radius $r_h$ is a universal horizon, and a boundary of the leaves of the foliation. Unlike GR, the surface $r=r_h$ is space-like, with time-like normal $u^N$. Indeed, it is none other than the surface at $t=\pm\infty$, as schematically argued in the introduction. Therefore, it contributes no additional boundary terms from Eq. \ref{bts}. 

The next step in transforming to a Hamiltonian is to write the Lagrangian in terms of $P_{MN}$, the momenta conjugate to the spatial metric of the ADM decomposition, $G_{MN}$\footnote{The purely spatial indices $I,J\dots$ can be extended to spacetime indices $M,N\dots$ through the definition of the spatial metric as a projector: $G_{MN}\equiv g_{MN}+u_Mu_N$.}. Recalling the definition of the extrinsic curvature in terms of the ADM fields, it is seen that only spatial derivatives of the shift $N_I$ appear in the action, highlighting its nature as a constraint. To transform to a Hamiltonian these spatial derivatives are integrated by parts to give another boundary term that contributes on the boundaries of the leaves of the foliation, $\d \Sigma_t$:
\begin{equation}
 -\frac{1}{8\pi G_K}\int_{\d \Sigma_t}\sqrt{H}d\Omega \frac{N^M P_{MN}n^N}{\sqrt{G}},
\end{equation}
where: $H$ is the determinant of the induced metric on surfaces of constant $t$ and $r$; $\Omega$ are the coordinates along these surfaces; the momenta is $P_{MN}\equiv\sqrt{G}(K_{MN}-\frac{1+c_2}{1-c_3}KG_{MN})$; and $n^N$ is the outward normal to the boundaries of the leaves of the foliation, $\d \Sigma_t$. 

The first contribution comes from the component of $\d \Sigma_t$ at the asymptotic boundary $r=0$. Here the outward normal $n^N$ is the space-like vector $s^N$, and this term generically contributes. The other possible contribution comes from the internal boundary of the universal horizon at $r=r_h$. Here the normal is $u^N$, the time-like vector that is orthogonal to the leaves of $\Sigma_t$, and therefore $P_{MN}u^N=0$ and there is no contribution. 

Putting this all together gives the value of the on-shell Hamiltonian for a solution to Ho\v rava gravity:
\begin{equation}\label{ham}
 \mathcal{H}\equiv -\frac{1}{8\pi G_K}\int_{S^0_t}\sqrt{H}d\Omega\left(\phantom{}^2\mathcal{K}N-\frac{N^IP_{IJ}s^J}{\sqrt{G}}\right),
\end{equation}
where $S^0_t$ is the surface at $r=0$ and constant $t$. For the asymptotically hyperbolic solutions at hand this quantity generically diverges. We therefore define the physical mass of a spacetime to be the difference between its on-shell Hamiltonian and that of a reference background. Importantly, as the Hamiltonians are regulated by a cut off near the $r=0$ boundary, this surface needs to be chosen for each background such that the value of the fields agrees there. Therefore the  lapse on the cut off is equal for each spacetime: $N(\epsilon)=N_0(\epsilon_0)$, where $N$ is the lapse for the solution under examination, evaluated on the surface $r=\epsilon$, while $N_0$ is the lapse of the reference background, evaluated on the surface $r=\epsilon_0$.

For $z=1$ the solution asymptotically approaches Anti-de Sitter space, which will be used as the reference background. Converting to the ADM coordinates of $\Sigma_t$, we can use the expansion Eq. \ref{zeq1exp} to determine the behavior of the integrands of Eq. \ref{ham} near $r=\epsilon$. For these solutions the first term concerning $\phantom{}^2\mathcal{K}$ contributes:
\begin{equation}
 -\frac{1}{8\pi G_K}\int_{S^\epsilon_t} d^2x\left(\frac{2}{\epsilon^3}+2C_e\right).
\end{equation}
The second term in the integrand involving the momenta $P_{IJ}$ contributes a term of order $\epsilon^3$, and therefore vanishes as the cut off is removed. Requiring that the lapse on the cut off surfaces of the solution and AdS space agree determines the relation:
\begin{equation}
 \frac{1}{\epsilon}+\frac{C_e\epsilon^2}{2}=\frac{1}{\epsilon_0}.
\end{equation}
Using Eq. \ref{ham} this determines the mass of the solution to  be:
\begin{equation}\label{mass}
 M_{z=1}\equiv\mathcal{H}_{z=1}-\mathcal{H}_{AdS}=\frac{-C_e A}{8\pi G_K},
\end{equation}
where $A\equiv \int d^2x$ is the volume of the transverse space.

For $z\ne 1$ one again finds that the on-shell Hamiltonian is divergent. In this case it can be regulated by performing a background subtraction by Lifshitz space, with appropriate $z$. Repeating the above arguments using the naive expansion Eq. \ref{zneq1exp} leads to the identical value Eq. \ref{mass} for the mass of the solution. On the other hand, the correct expansion for $z\ne 1$ has the subleading power $\Delta$ given in Eq. \ref{delta} which generically contributes to the mass. For the example given by the series Eq. \ref{z2delta2exp} the above procedure yields the mass:
\begin{equation}
 M_{z=2,\Delta=2}=\frac{(2C_a^2-C_e)A}{8\pi G_K}.
\end{equation}
The generic behavior of the mass of a $z\ne 1$ spacetime remains to be understood. An analytic solution going beyond an asymptotic series expansion would shed light on this point. Regardless, the definition of mass as the on-shell Hamiltonian Eq. \ref{ham} for a solution remains well defined, up to regularization as discussed. 

We could now derive a ``first law'' relating the variation of mass of two solutions that have a small variation of the dimensionful constants $C_e$ or $C_a$. As it stands this is not a very useful statement: from the asymptotic expansion alone there is no explicit relation between the parameters $C_e$ and $C_a$ and the radius of the universal horizon, $r_h$. A more rigorous  first law is generally derived by making use of the identity $\nabla^M(\nabla_N\chi_M)=R_{NM}\chi^M$, the equations of motion, and Gauss's law. Such a derivation can indeed be done in the case of Ho\v rava gravity and has been derived for Einstein-aether theory in \cite{Jishnu}. See also \cite{Mohd:2013zca} for a derivation following Wald's Noether charge method applicable to asymptotically flat solutions. An explicit example of the first law, following from an analytic solution, will be given in the next section.

\section{An analytic solution}\label{sec:anal}
For $z=1$, by examining the asymptotic expansion of the equations of motion Eq. \ref{zeq1exp} it is seen that the power series solutions for $e(r)$ and $f(r)$ appear to terminate. Therefore, making the ansatz:
\begin{equation}
 e(r)=\frac{1}{r^2}+C_e r-\frac{1}{4}c_3(C_e+2C_a)^2r^4,\qquad f(r)=\frac{1}{r^2},
\end{equation}
we see the equations of motion are solved by:
\begin{equation}
 a(r)=\frac{2\left(\sqrt{4+4C_er^3+(1-c_3)(C_e+2C_a)^2r^6}+(C_e+2C_a)r^3\right)}{4(1+C_er^3)r-c_3(C_e+2C_a)^2r^7}.
\end{equation}

This solution still depends on two parameters, $C_e$ and $C_a$, and it needs to be checked whether it is non-singular in the interior. As mentioned above, due to the nature of the equations of motion, a possible singular point of solutions is the sound horizon for the scalar mode with speed $s^2_0=((c_2+c_3)(2-c_4))/(c_4(1-c_3)(2+3c_2+c_3))$. For asymptotically AdS solutions, from Eq. \ref{asymplif}, $z=1$ implies that $c_4=0$, and therefore the scalar sound speed is $s_0\to \infty$. Intuitively, for an infinite speed scalar mode, its sound horizon should be at the same position as the universal horizon which traps modes of any speed. The previous Section \ref{sec:mass} determined the condition for the location of the universal horizon to be $\chi^M u_M=0$, for $\chi_M$ the asymptotically time-like Killing vector.

For the above analytic solution to be physical it must be non-singular at the universal horizon. One quantity to examine is $\left(\chi^M u_M\right)^2$: being a square it must be non-negative for a physical spacetime, while by above it must vanish at the universal horizon. Therefore its first derivative must also vanish there in order to satisfy these two properties. For the above solution:
\begin{equation}
 \left(\chi^M u_M\right)^2=\frac{1}{r^2}+C_er+(1-c_3)(C_e+2C_a)^2r^4,
\end{equation}
and requiring that this and its first derivative vanishes at the universal horizon, $r=r_h$, implies:
\begin{equation}
 C_e=-\frac{2}{r_h^3}\qquad C_a=\frac{1-1/\sqrt{1-c_3}}{r_h^3}.
\end{equation}
This simplifies the solution to be:
\begin{equation}\label{analsolu}
 e(r)=\frac{1}{r^2}-\frac{2r}{r_h^3}-\frac{c_3r^4}{(1-c_3)r_h^6},\qquad f(r)=\frac{1}{r^2},\qquad a(r)=\frac{r_h^3}{r_h^3r+\left(\frac{1}{\sqrt{1-c_3}}-1\right)r^4}.
\end{equation}

\subsection{Adapted coordinates and the universal horizon}\label{adptcoord}
It is important to note that the metric and foliation normal vector following from Eq. \ref{analsolu} are not written in the preferred global time of Ho\v rava gravity. In order to more fully understand the causal structure of this solution it is useful to change to the ADM coordinates adapted to the foliation. This is done by choosing the time coordinate to be the khronon $\phi$, so that $u_M$ has only a time component. For the above EF coordinates, $x^M=(v,r,\vec{x})$, and the ADM coordinates, $t$ and $y^I=(r,\vec{x})$, we have the following Jacobian's:
\begin{eqnarray}
 \nn t^M \equiv\frac{\d x^M}{\d t}, \qquad e^M_I\equiv\frac{\d x^M}{\d y^I},\\
\tilde{t}_M\equiv\frac{\d t}{\d x^M}, \qquad \tilde{e}^I_M\equiv\frac{\d y^I}{\d x^M}.
\end{eqnarray}

The global time of the ADM variables, $t$, and the null time of the EF coordinates $v$ are related by the ansatz $t=v+h(r)$, for $h$ a function of the radial coordinate $r$.  Under this coordinate change the hypersurface orthogonal vector $u_M$ becomes:
\begin{equation}\label{utrans}
 \tilde{u}_t= \frac{\d x^M}{\d t} u_M=u_v,\qquad \tilde{u}_I=e^M_I u_M=(-h'(r)u_v+u_r,\vec{0}).
\end{equation}
By definition of adapted coordinates we require $\tilde{u}_I=0$, and therefore determine $h'(r)=u_r/u_v$, which can be evaluated for the above solution. This then gives the ADM spatial metric $G_{IJ}$:
\begin{equation}
 G_{IJ}\equiv e^M_I g_{MN}e^N_J=\left(\begin{array}{ccc}
                                        h'(r)\left(g_{vv}h'(r)-2g_{vr}\right) & 0 & 0\\
0 & \frac{1}{r^2} & 0\\
0 & 0 & \frac{1}{r^2}
                                       \end{array}\right)=
\left(\begin{array}{ccc}
       \frac{r_h^6}{r^2(r^3-r_h^3)^2} & 0 & 0\\
	0 & \frac{1}{r^2} & 0\\
	0 & 0 & \frac{1}{r^2}
      \end{array}\right).
\end{equation}
The ADM shift vector $N_I$ is given by the time-space component of the transformed spacetime metric:
\begin{equation}\label{shift}
 N_I\equiv t^M g_{MN}e^N_I=\left(-h'(r)g_{vv}+g_{vr},\vec{0}\right)=\left(\frac{r}{\sqrt{1-c_3}(r_h^3-r^3)},\vec{0}\right).
\end{equation}
Lastly, the ADM lapse function $N$ is determined by the time-time component of the transformed spacetime metric:
\begin{equation}
 N^2\equiv N^IN_I-t^M g_{MN} t^{N}=\frac{(r^3-r_h^3)^2}{r^2r_h^6},
\end{equation}
giving the the final ADM form of the spacetime metric:
\begin{equation}\label{soluadm}
 ds^2=-\frac{(r^3-r_h^3)^2}{r^2r_h^6}dt^2+\frac{r_h^6}{r^2(r^3-r_h^3)^2}\left(dr-\frac{r^3(r^3-r_h^3)}{\sqrt{1-c_3}r_h^6}dt\right)^2+\frac{d\vec{x}^2}{r^2}.
\end{equation}
 
This allows one to interpret the meaning of the constant $r_h$ in the context of adapted coordinates. On the two-sphere at this radial coordinate the lapse function $N$ vanishes, but the lapse function is what tells us the normal distance from one spatial leaf of the foliation at global time $t_0$ to the next at $t_0+\Delta t$. Therefore the distance between the leaves is ever decreasing and they all bunch up as they approach $r=r_h$. As the preferred asymptotic global time runs to infinity, the leaves of $\Sigma_t$ never penetrate this two-sphere. This shows how causality and causal boundaries can arise in a theory like Ho\v rava gravity that have no intrinsic limiting speed. Disturbances can propagate arbitrarily fast, but they can only propagate forward in the preferred global time $t$. As the entire region exterior to $r=r_h$ is to the past of the interior region (always at an ``earlier'' global time) nothing can escape the interior. The location of the universal horizon is therefore a radius where the lapse function vanishes.

\subsection{Thermodynamics}
\subsubsection{Near horizon geometry and temperature}\label{nearhorgeo}
A classic method to derive the temperature of a black hole in GR is to examine the near horizon geometry. Generically the Euclidean manifold has a conical singularity at the Killing horizon unless the Euclidean time has a specific periodicity. The formalism of finite temperature QFT then dictates the temperature to be the inverse of this period. We therefore expect it to be beneficial to examine the geometry near the universal horizon, although our findings will be radically different then the generic case in GR.

Examining the near horizon $r\to r_h$ limit of the solution Eq. \ref{soluadm} can be tricky due to the off-diagonal $dtdr$ term coming from the shift vector Eq. \ref{shift}. In GR this term would be removed by a temporal diffeomorphism, but this is not allowed in the foliation preserving diffeomorphisms of Ho\v rava gravity. On the other hand, we are allowed to do a time-dependent radial diffeomorphism to eliminate this cross term. To this end it is first useful to redefine the radial coordinate:
\begin{equation}
 \rho\equiv\sqrt{1-c_3}r_h^6\left(\frac{1}{2r^2r_h^3}-ArcTan\left(\frac{2r+r_h}{\sqrt{3}r_h}\right)+\frac{Log(r_h-r)}{3r_h^5}-\frac{Log(r^2+r_hr+r_h^2)}{6 r_h^5}\right),
\end{equation}
which goes to $+\infty$ at the boundary $r\to 0$, and $-\infty$ at the universal horizon $r\to r_h$. The reason for this definition is that it makes the $dr+N^rdt$ term of Eq. \ref{soluadm} conformally flat. We then can define a new time dependent radial coordinate $\xi=\rho-t$ which diagonalizes the metric. For a fixed $t$, $\xi$ has the same behavior as $\rho$: $\xi\to+\infty$ at the boundary, and $\xi\to-\infty$ at the horizon. This has the price of making the metric non-static, giving the near-horizon behavior:
\begin{equation}
 r_h-r\approx exp(3\rho/(\sqrt{1-c_3}r_h))=exp(3(\xi+t)/(\sqrt{1-c_3}r_h)),
\end{equation}
which gives the near-horizon form of the metric:
\begin{equation}
 ds^2\approx-\frac{9e^\frac{6(\xi+t)}{\sqrt{1-c_3}r_h}dt^2}{r_h^4}+\frac{d\xi^2}{r_h^2(1-c_3)}+\frac{dx^2}{r_h^2}.
\end{equation}
The non-static term can be removed by performing an allowed time reparametrization: 
\begin{equation}
 \label{tau} \tau=\frac{\sqrt{1-c_3}r_h exp(3t/(\sqrt{1-c_3}r_h))}{3},
\end{equation}
which goes to $0$ as $t\to -\infty$, and $\tau\to +\infty$ as $t \to +\infty$. One last radial coordinate $R$ can be defined to put the $\xi$-$\tau$ terms of the metric into a conformal form:
\begin{equation}
 \frac{d\xi^2}{r_h^2(1-c_3)}\equiv\frac{9 exp(6\xi/(\sqrt{1-c_3}r_h)dR^2}{r_h^4}.
\end{equation}
This gives the final form of the near-horizon metric:
\begin{equation}
 ds^2=\frac{1}{(3R)^2}\left(-d\tau^2+dR^2\right)+\frac{dx^2}{r_h^2},
\end{equation}
which is recognized to be $AdS_2$ with curvature radius $1/3$ (relative to the asymptotic $AdS_4$ geometry of the solution) crossed with $\mathbb{R}^2$.

Unlike generic black holes in GR, the near horizon geometry is not Rindler space. Therefore an approach following Euclideanization and the subsequent periodicity of imaginary time is not available to define a temperature. Fortunately, despite having no extrinsic scale, it is understood how $AdS_2$ can have a notion of temperature \cite{Spradlin:1999bn}. This arises because different choices of time coordinate in $AdS_2$ can lead to inequivalent vacua of a QFT upon quantization. For the case at hand, there is an inherited time $t$, which is the Poincar\'e time of the decoupled  asymptotic $AdS_4$ geometry. This is the time that would correspond to the Minkowski time of the dual NR QFT in flat space. Of interest is the behavior of objects like correlation functions with respect to the time $t$. Due to the relation between the $AdS_2$ time $\tau$ and the $AdS_4$ time $t$ from Eq. \ref{tau}, it is seen that the former is periodic in the imaginary part of the latter. This implies that any calculation performed in the vacuum of the near horizon geometry will be periodic in the imaginary part of $t$, and therefore exhibits thermal behavior from the view of the boundary observer. The inverse of this period gives the temperature of the spacetime:
\begin{equation}\label{temp}
 T_H=\frac{3}{2\pi r_h\sqrt{1-c_3}}.
\end{equation}

\subsubsection{Entropy and the first law}
The formula Eq. \ref{mass} for the mass of an asymptotically hyperbolic spacetime allows the first law to be displayed for this black hole. The solution Eq. \ref{soluadm} has $C_e=-2/r_h^3$ and therefore determines the mass to be:
\begin{equation}
 M=\frac{A}{4\pi G_K r_h^3}.
\end{equation}
Taking the on-shell action to be the Helmholtz free energy divided by temperature allows the calculation of the entropy. The on-shell action needs to be regulated by a background subtraction, and the extent of temporal integration of the two spacetimes is related to maintain the same geometries on the cut off surfaces \cite{Witten:1998zw}. This process yields the regulated on-shell action $I$ for our solution:
\begin{equation}
I=\frac{-\beta A}{8\pi G_K r_h^3}, 
\end{equation}
where $\beta$ is the inverse of the black hole temperature $T_H$ Eq. \ref{temp}. Thermodynamic identities now give the entropy:
\begin{equation}\label{entropy}
S\equiv \beta M - I = \frac{3\beta A}{8 \pi G_K r_h^3}= \frac{\sqrt{1-c_3}A_h}{4G_N},
\end{equation}
where $A_h$ is the transverse area of the horizon and, recalling Eq. \ref{newtonconst}, for this solution $G_N=(1-c_3)G_H=G_K$ is the Newton constant. The first law is thus checked and this Ho\v rava black hole is seen to obey sensible thermodynamics.  

\subsubsection{Tunneling method}
An alternate and intuitively pleasing method to calculate a black hole's temperature is the so-called tunneling method \cite{Parikh:1999mf}. The foundational idea is that near a horizon the virtual pairs of particles in the quantum vacuum can be disassociated with the ``negative mass'' partner ending up in the black hole, in order to maintain energy conservation. The positive energy partner is then free to travel to the asymptotic region and be interpreted as Hawking radiation.

Calculationally, this method makes use of the fact that a given quanta of this radiation was extremely blueshifted near the horizon, and therefore a semiclassical approach can be used. Considering a scalar field, this allows a WKB approximation of the wavefunction $\Phi$ of an excitation as:
\begin{equation}\label{wkb}
 \Phi(x)\approx \phi_0 exp\left(\imath \mathcal{S}[\phi(x)]\right),
\end{equation}
where $\mathcal{S}$ is the action of the scalar field, $\phi$ is its classical solution, and $\phi_0$ is a constant amplitude. This wavefunction determines the rate that the scalar field can tunnel through the horizon. With the WKB approximation the quantum probability of emission is:
\begin{equation}
 \Gamma\equiv \Phi^*\Phi\propto exp(-2 \ Im[\mathcal{S}]).
\end{equation}
If this emission distribution is Boltzmann than a temperature can be meaningfully associated to the process.

Although the WKB approximation requires knowledge of the equation of motion of the scalar field, the simpler eikonal/Hamilton-Jacobi approximation only requires a dispersion relation for the field. From Eq. \ref{wkb} the Hamilton-Jacobi equations can be derived by acting the momentum operator on the wavefunction $\Phi$:
\begin{equation}
 k_M=\nabla_M\mathcal{S}[\phi(x)].
\end{equation}
Combining this with the dispersion relation for $\phi$ allows the determination of $Im[\mathcal{S}]$, and therefore the tunneling probability.

In GR it is argued that due to the extreme blueshift near the horizon the appropriate dispersion relation to use is that of a massless particle, $k_t^2=\vec{k}^2$, regardless of the actual equation of motion of the field. While this dispersion relation is fixed in GR due to Lorentz invariance, in Ho\v rava gravity it could take on the very general form:
\begin{equation}\label{dispersion}
 k_t^2= \frac{\vec{k}^{2z_\phi}}{k_0^{2(z_\phi-1)}} +\cdots,
\end{equation}
where $z_\phi$ is an integer determining the nature of the dispersion, $k_0$ is a constant of dimension one, and the dots schematically imply that all powers of $\vec{k}^2$ less than $z_\phi$, as well as derivatives of $\vec{k}$ can be included, see \cite{Jishnu} for a more precise discussion.

A tractable, and seemingly natural, choice is $z_\phi =2$. This gives a dispersion that is similar to the Schr\"odinger equation. Proceeding with the traditional methods of calculating $Im\mathcal{S}$, one obtains the temperature of a static spherically symmetric black hole in Ho\v rava gravity \cite{Jishnu,Berglund:2012fk}:
\begin{equation}
 T_{UH}=\left.\frac{a^Ms_M|\chi|}{4\pi}\right|_{r=r_h},
\end{equation}
where $s_M$ is the outward pointing space-like vector orthogonal to $u_M$, $a_M\equiv u^N\nabla_N u_M$, and $\chi^M=(1,0,0,0)$ is the asymptotically time-like Killing vector. For the above analytic solution Eq. \ref{soluadm} this gives the temperature:
\begin{equation}
 T_{UH}=\frac{1}{2}T_H,
\end{equation}
that is, one half of the value as determined by the geometric method of Section \ref{nearhorgeo}. Interestingly, another tractable value is the $z_\phi\to\infty$ limit \cite{Jishnu}, for which the temperature is twice that of the $z_\phi=2$ case, and therefore agrees with the geometric method.

The fact that the tunneling method calculation has the ambiguity of the free parameter $z_\phi$ makes it a somewhat unappealing technique to calculate the Hawking temperature, which is expected to be universal. In GR, the possible ambiguity arising for fields of different mass or spin has been shown to be irrelevant in the calculation of Hawking radiation because the extreme blueshift of the near-horizon region dictates that only the linear high energy dispersion relation, constrained by Lorentz invariance, plays a role. It can be hoped that similarly the high energy dispersion relation of fields near the universal horizon in Ho\v rava gravity is unique. Some evidence, presented in Section \ref{sec:conc}, points towards the $z_\phi=\infty$ limit being this unique relation. In this case the temperature as calculated via the tunneling method agrees with that extracted from the near horizon geometry.

The temperature derived via the tunneling method for $z_\phi=2$ has recently been reproduced in \cite{Cropp:2013sea} by arguments concerning ray tracing near the universal horizon and their relation to the surface gravity. It is claimed this calculation is universal regardless of the exact dispersion, but that would contradict the tunneling method in the $z_\phi\to\infty$ limit. Whether the ray tracing method can be used in this regime would provide crucial insight into the temperature of universal horizons. 

\subsection{An asymptotically flat solution}
Another analytic black hole which provides a testing ground of the ideas developed above has been presented in \cite{Jishnu}. The four dimensional metric is given as:
\begin{equation}\label{flatbh}
 ds^2=-e(r)^2dt^2+\frac{1}{e(r)^2}\left(dr-f(r)e(r)dt\right)^2+r^2 d\Omega^2,
\end{equation}
where:
\begin{equation}
 e(r)=-1+\frac{r_h}{r},\qquad f(r)=\sqrt{\frac{\mu r_h}{r}+\frac{(2-c_4)r_h^2}{2(1-c_3)r^2}}.
\end{equation}
Asymptotic infinity is the region $r\to \infty$, which is seen to be Minkowski space in spherical coordinates. Importantly, the metric is written in a preferred global time, and therefore $e(r)$ is the lapse function of the ADM decomposition and, as above, it vanishing at $r=r_h$ signals the location of a universal horizon. The constant $\mu$ further parametrizes the solution.

Similar manipulations as applied to the asymptotically AdS black hole of Section \ref{nearhorgeo} can be brought to bear. After diagonalizing the metric by performing a time-dependent radial diffeomorphism, followed by a time reparametrization, the near horizon geometry of Eq. \ref{flatbh} becomes:
\begin{equation}
 ds^2\approx \frac{r^2_h}{R^2}\left(-d\tau^2+dR^2\right)+r_h^2d\Omega^2,
\end{equation}
where: $R\to \infty$ at the universal horizon; and we again recognize the geometry to be $AdS_2$, now crossed with a sphere, both of radius $r_h$.

The temperature of this solution follows from the relation between the near horizon $AdS_2$ time $\tau$ and the asymptotic Minkowskian time $t$:
\begin{equation}
 \tau\equiv-\frac{r_h}{\sqrt{\mu +\frac{2-c_4}{2(1-c_3)}}} exp\left(-\frac{r_h t}{\sqrt{\mu +\frac{2-c_4}{2(1-c_3)}}}\right).
\end{equation}
Periodicity in the imaginary part of $t$ determines the temperature to be:
\begin{equation}
 T_H=\frac{1}{2\pi r_h}\sqrt{\mu +\frac{2-c_4}{2(1-c_3)}}.
\end{equation}
Comparing to the calculation of temperature presented in \cite{Jishnu}, we again see that the tunneling method agrees with this geometric method for $z_\phi=\infty$, and not for $z_\phi=2$. 

\section{Numerical solutions}\label{sec:num}
\subsection{Probe limit}
As discussed in Section \ref{khronon}, a powerful use of the khronon formalism is the probe limit regime. When the khronon does not backreact, any solution to GR is a solution to Ho\v rava gravity; the khronon simply determines what time coordinate needs to be used to be a preferred global time. 

An interesting class of manifolds to examine with this technique are those which are black holes of GR \cite{Blas:2011ni}. These are defined by having event horizons: null causal boundaries of the asymptotic region. Whether the event horizon\footnote{From hereon the event horizon will be referred to as the metric horizon to avoid confusion with the universal horizon, which is the true causal boundary of Ho\v rava gravity.} maintains an important status can be explored with the probe limit technique. 
A particularly interesting black hole geometry, from the viewpoint of holography, is the Anti-de Sitter-Schwarzschild (AdS-S) solution. In four dimensions the metric in Poincar\'e coordinates is:
\begin{equation}
 ds^2_{AdS-S}=-\frac{1-r^3}{r^2}dt^2+\frac{1}{r^2(1-r^3)}dr^2+\frac{1}{r^2}d\vec{x}^2,
\end{equation}
where the boundary is at $r=0$, and the radius of the metric horizon has been set to one. It is beneficial to work in coordinates that are non-singular at the metric horizon. Using the Eddington-Finkelstein time $v\equiv t+r_*$, where the tortoise coordinate is given by $r_*=\int^rdr'/(1-r'^3)$, the metric becomes:
\begin{equation}
 ds^2_{AdS-S}=-\frac{1-r^3}{r^2}dv^2+\frac{2}{r^2}dvdr+\frac{1}{r^2}d\vec{x}^2.
\end{equation}

On this background the khronon equation of motion can be derived from the probe action\footnote{This is because $R_{MN}u^Mu^N=const$ for AdS-S and a normalized $u_M$.}:
\begin{equation}\label{probeaction}
 S_\phi=-\frac{c_4}{16\pi G_K}\int dv dr dx^2\sqrt{-g}\left[\frac{1}{2}F_{MN}F^{MN}+s_0^2(\nabla_M u^M)^2\right],
\end{equation}
recalling that $F_{MN}\equiv \d_M u_N-\d_N u_M$, and $s_0$ is the sound speed of the scalar mode of Ho\v rava gravity, given by $s_0^2=(c_2+c_3)/c_4$ in the probe limit, whereas the sound speed of the spin two graviton is $s_2= 1$. From the action Eq. \ref{probeaction}, it is seen that in the probe limit the only effective coupling constant is the speed $s_0$. A useful parametrization of the khronon orthogonal vector for a static, transversely constant ansatz is:
\begin{equation}\label{eq:u}
 u_M=\left(-\frac{1+f(r)}{2r}\sqrt{\frac{1-r^3}{f(r)}},\frac{1}{r}\sqrt{\frac{f(r)}{1-r^3}},0,0\right).
\end{equation}
As such it is explicitly normalized to be unit time-like in the AdS-S background, $u_M u^M=-1$. The boundary condition that the global time is asymptotically Poincar\'e time requires $\phi\to t=v+r_*$ at the boundary. In EF coordinates this is equivalent to $u_M\to(-1/r,1/r,0,0)$, or $f(0)=1$ for the above parametrization.

Varying the action Eq. \ref{probeaction} with respect to $f(r)$ gives the equation of motion for the khronon. This second order non-linear ODE is seen to have the expected singular points at the boundary, $r=0$, and the metric horizon, $r=1$. Additionally, there is a singular point whenever the magnitude of the khronon vanishes, $f(r_{crit})=0$, but, from examining the expression Eq. \ref{eq:u} for the khronon normal vector, $u_M$ is non-singular for $f(r_{crit})=0$ only if $r_{crit}=1$, that is, this is not a new singular point, but just the metric horizon again. This is understood by recognizing that in the probe limit $s_2=1$ and therefore the metric horizon is more properly understood as the sound horizon for the spin two degrees of freedom of the metric. Lastly, there is a singular point at the sound horizon for the scalar mode, $r_s$, where the magnitude of the khronon function satisfies
\begin{equation}\label{fs}
 f(r_s)^2-2f(r_s)\frac{1+s_0^2}{1-s_0^2}+1=0.
\end{equation}
This gives $f_s\equiv f(r_s)=(1\pm s_0)^2/((1-s_0)(1+s_0))$. The bottom sign is chosen as the physically acceptable condition for two reasons: there should be a non-singular solution for the limit $s_0\to1$ corresponding to the scalar and spin two sound horizons coinciding, which requires $f\to 0$ as above; additionally, in the following numerical construction, solutions using the top sign either do not match the desired boundary condition $f(0)=1$ or are singular at the metric horizon.

A final important radial coordinate can be seen from the parametrization for $u_M$, Eq. \ref{eq:u}. Recall from Section \ref{khronon} that when written in the preferred global time the hypersurface normal vector $u_M$ has only a temporal component which is given by the ADM lapse function $N$. Additionally, in transforming from the EF time $v$ to the preferred global time the temporal component of $u_M$ is unaltered, see Eq. \ref{utrans}. This implies that at a coordinate where $u_v$ vanishes, so does the lapse function $N$ of the preferred foliation and therefore this is the location of the universal horizon, $r_h$. From the parametrization Eq. \ref{eq:u} this determines the universal horizon to be the radial coordinate such that $f(r_h)=-1$. This further implies, from examining $u_r$, that physical solutions must have $r_h>1$, that is the universal horizon is inside of the metric horizon. 

\subsection{Solutions}
The second order non-linear ODE for $f(r)$ can be numerically solved via a shooting method. At the scalar sound horizon the function is given by $f(r_s)=f_s$, see Eq. \ref{fs} above. Requiring a regular solution at this singular point in turn determines the value of $f'(r_s)$ from two possible choices. Using this data as boundary conditions, numerical integration can be performed in either direction, increasing or decreasing $r$, to find a full solution. In practice, we instead impose boundary conditions a small distance $\epsilon$ from the scalar sound horizon, and use a Taylor series that matches the desired behavior at $r_s$. This is the typical trick to improve numerical stability while integrating near a singular point. The technique of matching a numerical solution to a local Taylor series must also be done to jump over the singular point at the metric horizon $r=1$.

For a given scalar speed $s_0$, taking the location of the sound horizon $r_s$ as the shooting parameter, and using the previously mentioned boundary conditions there, the equation of motion can be numerically integrated out to the boundary $r=0$. By varying $r_s$ one can obtain the value that is needed to meet the boundary condition $f(0)=1$, which is the requirement that the foliation asymptotically tends to that of Poincar\'e time. For every $s_0$ this gives a unique regular solution that is asymptotically AdS.

\subsubsection{Case I: $s_0<1$}
\begin{figure}[!htb]
\centering
\includegraphics[width=0.62\textwidth]{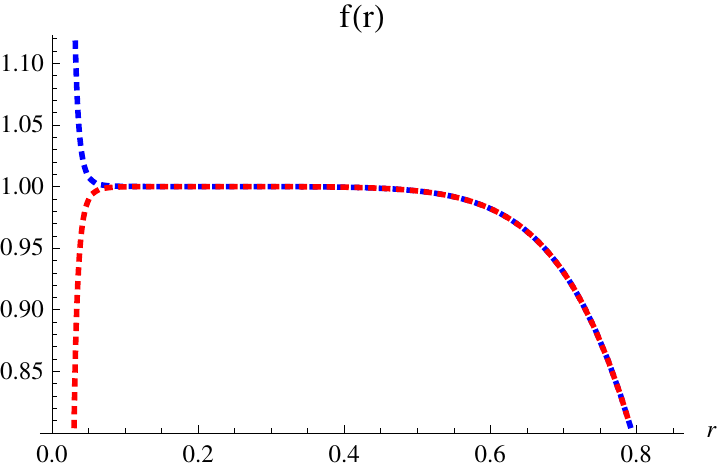}
\caption{The profile of the khronon function $f(r)$ for $s_0=2/10$ is plotted from $r_s$ to the boundary at $r=0$. The lower red branch has $r_s=212192121/250000000\sim 0.848768$ while the upper blue branch has $r_s$ larger by $10^{-9}$.}
\label{fig:fig1}
\end{figure}

For the scalar speed $s_0<1$ the sound horizon is outside of the metric horizon, that is $r_s<1$. To implement the above numerical procedure we use a Taylor series expansion about $r_s$ which solves the equation of motion to order $(r-r_s)^4$ and implement the required boundary conditions at $r=r_s-\epsilon$ for $\epsilon=10^{-3}$. Figure \ref{fig:fig1}
shows two plotted solutions for $r_s$ differing by $10^{-9}$, the accuracy used throughout. The desired solution with $f(0)=1$ lies between the plotted two. Figure \ref{fig:fig2}
\begin{figure}[!htb]
\centering
\begin{tabular}{c|c|c}
$s_0$ & $r_s$ & $r_h$\\
\hline
\hline 0.1 & 0.823037721 & 1.12386952\\
\hline 0.2 & 0.848768484 & 1.13782822\\
\hline 0.3 & 0.872378252 & 1.15121426\\
\hline 0.4 & 0.894577562 & 1.16436206\\
\hline 0.5 & 0.915465270 & 1.17473414\\
\hline 0.6 & 0.935021618 & 1.18484700\\
\hline 0.7 & 0.953227190 & 1.19334154\\
\hline 0.8 & 0.970092978 & 1.20092572\\
\hline 0.9 & 0.985661688 & 1.20752533\\
\end{tabular}
\caption{The location of the scalar sound horizon $r_s$, and the radius of the universal horizon $r_h$, for speeds $s_0<1$.}
\label{fig:fig2}
\end{figure}
presents crucial results for speeds $s_0<1$: the $r_s$ giving the correct boundary conditions is shown, along with $r_h$, the radial coordinate at which $f(r)$ first becomes equal to $-1$, which is the location of the universal horizon, as argued above. 

\begin{figure}[!htb]
\centering
\includegraphics[width=0.62\textwidth]{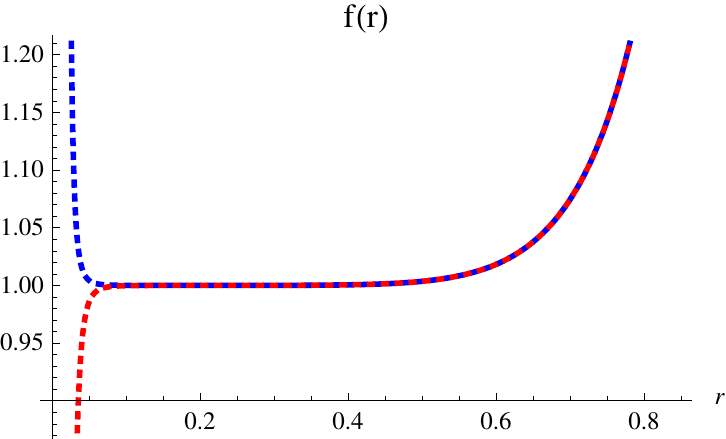}
\caption{The profile of the khronon function $f(r)$ for $s_0=2/10$ with incorrect boundary conditions at $r_s$. Although the behavior is as desired at $r=0$, $f(r)$ is divergent at the metric horizon $r=1$.}
\label{fig:fig3}
\end{figure}

We reiterate that the three other possible combinations of boundary conditions for $f(r_s)$ and $f'(r_s)$ do not give physically acceptable solutions. Two of them always have $f(0)>1$ or $f(0)<1$ for all $r_s$, never switching as in Figure \ref{fig:fig1}. This means they do not realize the desired condition of Poincar\'e time at the asymptotic boundary. The final possible class of solutions, those with $f_s=(1+s_0)/(1-s_0)$ and $f'(r_s)>0$, do exhibit the correct asymptotic $f(0)\to1$ behavior, but are divergent at the metric horizon. Examining $u_M$ in Eq. \ref{eq:u}  we see that this leads to a non regular solution for the hypersurface orthogonal vector. An example of this class is shown in Figure \ref{fig:fig3}.

\subsubsection{Case II: $s_0>1$}
For the khronon speed $s_0>1$ the scalar sound horizon is inside the metric horizon, that is $r_s>1$. The numerics are implemented by the shooting method as above. Figure \ref{fig:fig4} gives the data of the solutions: $r_s$ and $r_h$ for each $s_0$.
\begin{figure}[!htb]
\centering
\begin{tabular}{c|c|c}
$s_0$ & $r_s$ & $r_h$\\
\hline
\hline 1.1 & 1.013189612 & 1.218154746\\
\hline 1.2 & 1.025420211 & 1.223152116\\
\hline 1.3 & 1.036500000 & 1.226276947\\
\hline 1.4 & 1.046758982 & 1.229434208\\
\hline 1.5 & 1.056222636 & 1.232232389\\
\hline 1.6 & 1.064967826 & 1.234712679\\
\hline 1.7 & 1.073108780 & 1.237040172\\
\hline 1.8 & 1.080589512 & 1.238919367\\
\hline 1.9 & 1.087514072 & 1.240540799\\
\hline 2   & 1.094025032 & 1.242147735\\
\hline 4   & 1.166242711 & 1.254600129\\
\hline 8   & 1.210285515 & 1.258505374\\
\hline 16  & 1.234393424 & 1.259559808\\
\hline 32  & 1.246977962 & 1.259830158\\
\hline 64  & 1.253404370 & 1.259898282\\
\hline 128 & 1.256651377 & 1.259915354
\end{tabular}
\caption{The location of the scalar sound horizon $r_s$, and the radius of the universal horizon $r_h$, for speeds $s_0>1$.}
\label{fig:fig4}
\end{figure}
The dependence of $r_s$ and $r_h$ on the speed $s_0$ is show in Figure \ref{fig:fig5}. 
\begin{figure}[!htb]
\centering
\includegraphics[width=0.62\textwidth]{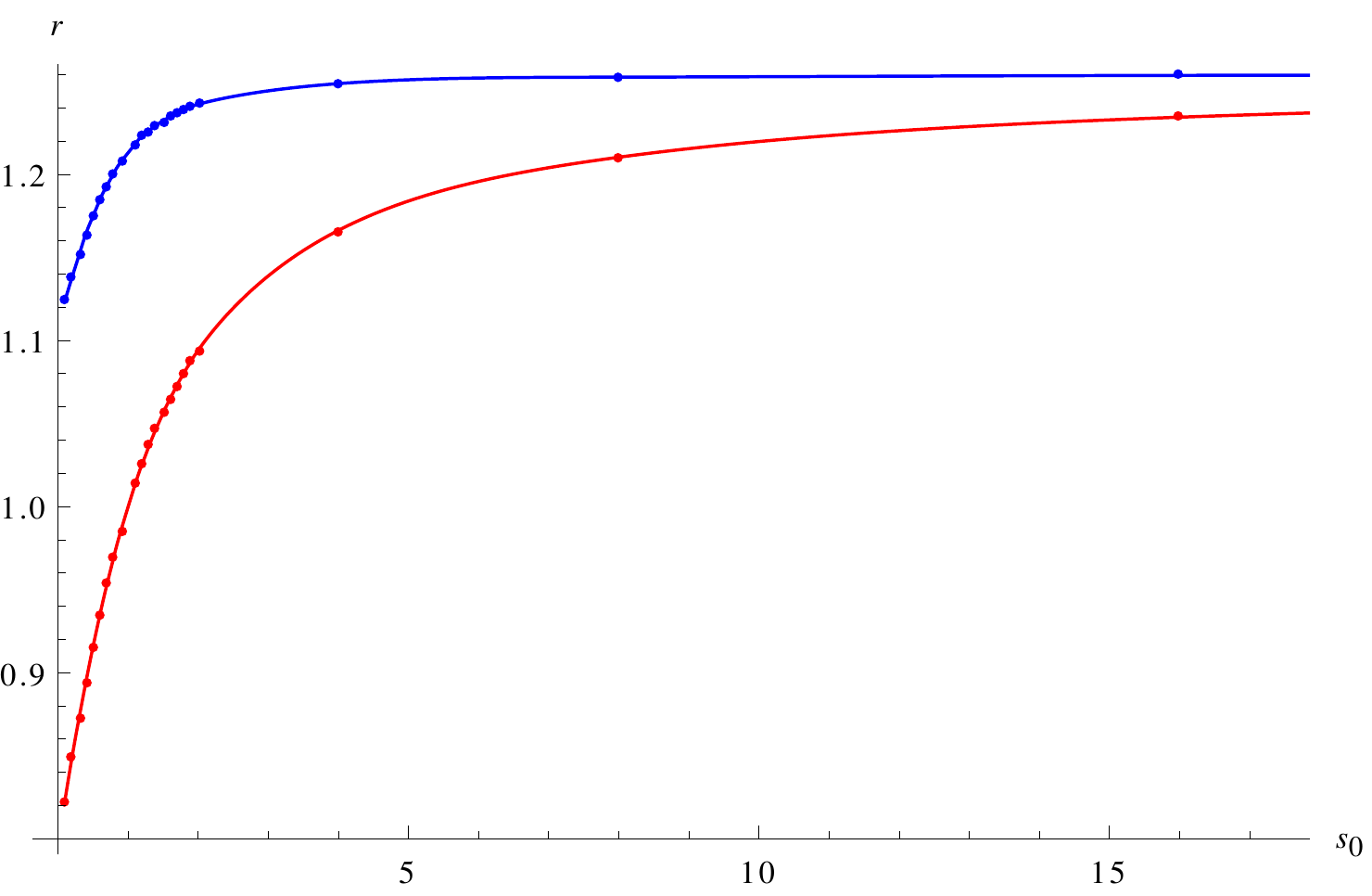}
\caption{The dependence of the scalar sound horizon $r_s$ on the speed $s_0$ (lower, red), and the dependence of the universal horizon $r_h$ (upper, blue).}
\label{fig:fig5}
\end{figure}
Of importance is that the universal horizon $r_h$ is always behind the metric horizon at $r=1$, and always behind the scalar sound horizon $r_s$. From the data it appears that $r_h$ is bound between $(1.11,1.26)$ as $s_0\to(0,\infty)$, respectively. In fact, the analytic solution of Section \ref{sec:anal} can be used to determine the $s_0\to\infty$ behavior. Since this solution is asymptotically Anti-de Sitter, if we take the probe limit it will correspond to one of the numerical solutions above. Recalling that the solution has $c_4=0$, in the probe limit this solution has a scalar speed of $s_0^2=(c_2+c_3)/c_4\to \infty$. From Eq. \ref{analsolu}, in the probe limit, the analytic solution has the metric component:
\begin{equation}
 g_{vv}= -\frac{1}{r^2}+\frac{2r}{r_h^3}.
\end{equation}
The metric horizon is where this component vanishes, giving $g_{vv}=-1+2/r_h^3=0$, that is $r_h=2^{1/3}\approx1.2599$, agreeing with the numerical bound above.

\subsection{Universal horizon}
\begin{figure}[!htb]
\centering
\includegraphics[width=0.62\textwidth]{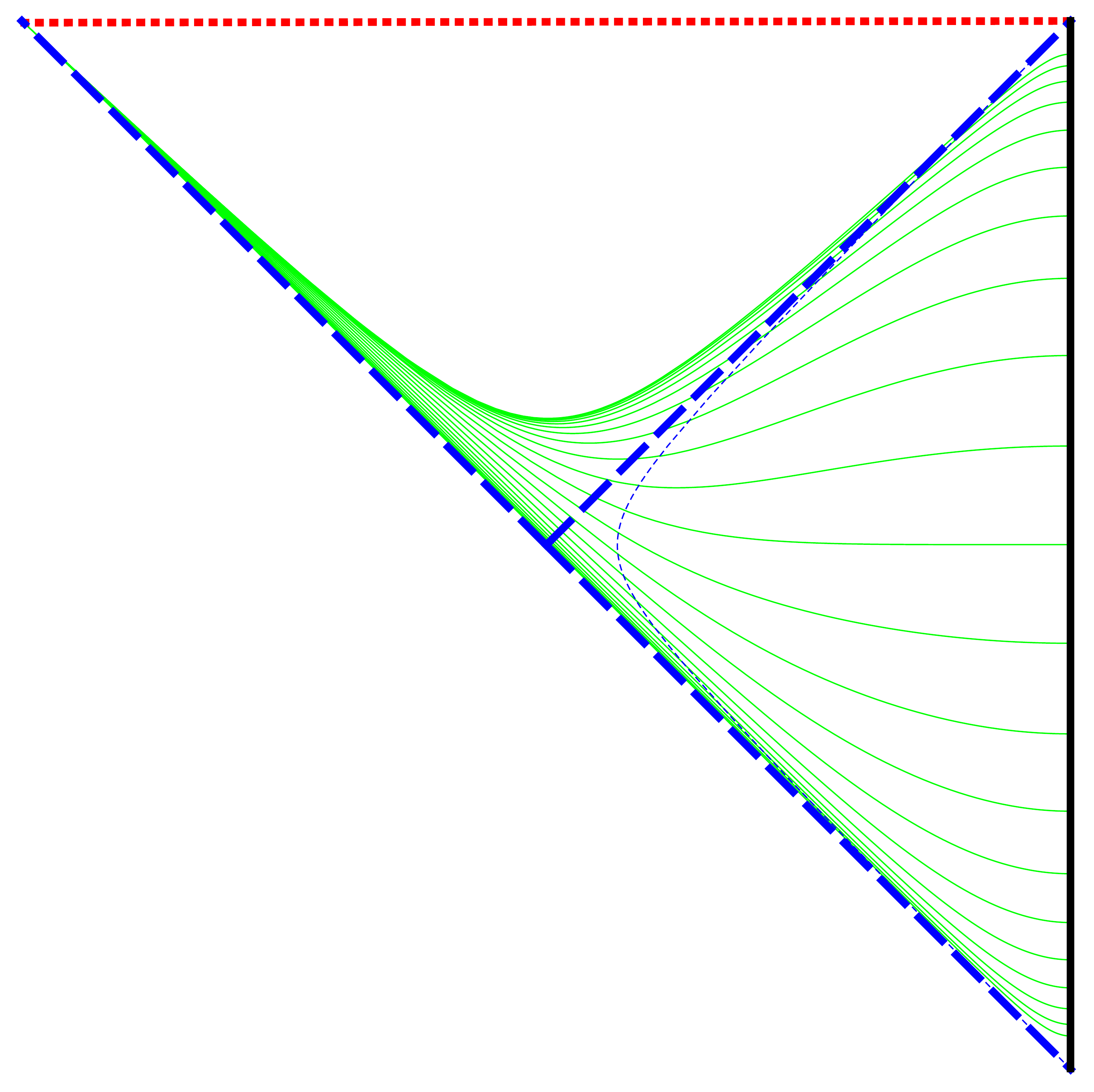}
\caption{The Penrose diagram of the AdS black hole. The singularity at $r\to\infty$ is in dotted red at the top, the boundary $r=0$ is in thick black on the right, the metric horizon at $r=1$ is in thick dashed blue, while the thin dashed blue is the scalar sound horizon at $r_s$. The level sets of the khronon with $s_0=7/10$ are shown in thin green.}
\label{fig:fig6}
\end{figure}

These numerical solutions all have the behavior that $u_v\to0$ as $r\to r_h$. By writing the khronon as $\phi=v+h(r)$ it is easy to see that $h(r)=\int^r dr' u_r/u_v+const$, as in Section \ref{adptcoord}. Therefore the vanishing of $u_v$ implies that the khronon diverges\footnote{This divergence is physically acceptable as it can be removed via an allowed temporal reparametrization. The invariant field $u_M$ is non singular.} at this radius, consequently the spatial slices of the foliation defined by the level sets of $\phi$ pile up at this radius and do not penetrate to smaller $r$. This is shown, for $s_0=7/10$, in the Penrose diagram of  Figure \ref{fig:pensads}, repeated here in Figure \ref{fig:fig6} for clarity.
Despite foliating the entire exterior of the black hole the leaves coming from the asymptotic boundary pile up at $r_h\approx 1.19$ and do not reach further into the interior. It is important to note that the interior region $r>r_h$ still has a foliation by a preferred global time: Figure \ref{fig:fig6} only shows the foliation that is connected to the asymptotic boundary for clarity, the numerical solution for the khronon, and hence the foliation, can be constructed arbitrarily far into the interior, as in \cite{Barausse:2011pu}. This shows that the foliations of the interior are disconnected from the foliation that reaches the asymptotic boundary. 

The khronon defines the preferred global time coordinate of Ho\v rava gravity, while causality means that influences can only propagate forward in global time, although arbitrary speed is allowed. In particular any disturbance at $r>r_h$ can only propagate towards larger $r$: everywhere exterior to $r_h$ is at an ``earlier'' global time. In this sense the surface $r_h$, which is the boundary of the exterior foliation, defines a causal boundary for the asymptotic observer and is therefore justly a universal horizon. 

The status of the metric horizon at $r=1$ can now be made clear. In Ho\v rava gravity this surface is properly understood as the sound horizon for the spin 2 graviton. Like the scalar sound horizon at $r_s$, these spheres are trapped surfaces for the low energy modes of their respective gravitons. On the other hand, higher energy corrections to Ho\v rava gravity will modify the dispersion relations of the gravitons to allow arbitrary speed\footnote{These corrections are not expected to modify the low energy picture of the universal horizon as the curvature is generically small there.}. This allows the gravitons, and any other fields considered, to escape their respective sound horizons, but they will still be inevitably trapped by the universal horizon as this is a trapped surface for modes of infinite speed. Therefore the relative unimportance of the metric horizon as a sound horizon, as compared to the universal horizon as a causal boundary, is understood in Ho\v rava gravity.

\section{Discussion}\label{sec:conc}
Herein the class of solutions to Ho\v rava gravity that asymptote to Lifshitz spacetimes are explored. These are an important class of solutions from the viewpoint of holographic application. Lifshitz spacetimes in GR have been argued to be dual to non-relativistic quantum field theories. Given that these spacetimes are vacuum solutions of Ho\v rava gravity lends evidence to the proposal that this alternate theory of gravity is dual to generic NR QFTs \cite{Janiszewski:2012nb,Janiszewski:2012nf,Griffin:2012qx}.

Very important to holography are black hole spacetimes and the thermodynamics they obey. The main result of this paper is that the analytic black hole of Section \ref{sec:anal} has both a temperature and an entropy, and that it obeys a first law of thermodynamics. Although this is very reassuring, the calculations of Section \ref{sec:anal} raised some interesting questions. 

The first of these is that the near horizon geometry is not Rindler space, as it generically is in GR. This also arises in the asymptotically flat black hole of \cite{Jishnu}. Instead the geometry is $AdS_2$ crossed with the transverse space, $\mathbb{R}^2$ or $S^2$, respectively. This indeed appears to be the generic situation in Ho\v rava gravity: the universal horizon occurs where the lapse $N$ vanishes, if it does so linearly, then $g_{tt}$ vanishes quadratically, once the metric is diagonalized as in Section \ref{nearhorgeo}. Despite this, a notion of temperature can still be defined via the methods of \cite{Spradlin:1999bn}.

The second interesting, and likely related issue, arises in the calculation of the temperature via the tunneling method. This approach has the weakness that the dispersion relation Eq. \ref{dispersion} is not unique once Lorentz invariance is abandoned. When using the tunneling method in GR the extreme blueshift of the horizon is used to justify the linear dispersion $k_t=|\vec{k}|$ regardless of the mass of the particle. In Ho\v rava gravity the same logic can be used. For modes of extremely high momenta, the leading power $z_\phi$ in Eq. \ref{dispersion} dominates all lower powers in the dispersion. As $z_\phi$ is not uniquely fixed by symmetry, as it is in GR, this argument favors as large a $z_\phi$ as possible. It therefore seems natural to use $z_\phi\to\infty$ for the dispersion in the near universal horizon calculation of the tunneling method. It is also reassuring that this form of the dispersion gives a temperature that agrees with the geometric method of Section \ref{nearhorgeo}, while the ``minimal'' choice $z_\phi=2$ disagrees by a factor of two. Why this may be related to the near universal horizon geometry is the fact that $AdS_2$ can be seen as the $z\to\infty$ limit of the Lifshitz spacetime Eq. \ref{eq:lif}. A better understanding of the natural action for a non-relativistic scalar in these background geometries will prove crucial in justifying these arguments.    

\section*{Acknowledgements}
I would like to thank Andreas Karch for his great insight and far reaching knowledge of the literature, without his guidance this paper could not exist. Jishnu Bhattacharyya, Matthew Roberts, Brandon Robinson, and David Sommer all contributed crucially to my understanding of this topic. This work was supported in part by the U.S. Department of Energy under Grant No. DE-FG02-96ER40956.     

\bibliographystyle{JHEP}
\bibliography{horsol}

\end{document}